\documentclass[aps,prc,a4paper,nofootinbib,preprintnumbers,superscriptaddress,showpacs,showkeys,twocolumn]{revtex4}

\usepackage{amsmath}
\usepackage{amssymb}
\usepackage{bm}
\usepackage{graphicx}
\usepackage{color}
\usepackage[english]{babel}

\newcommand{\be}{\begin{equation}}
\newcommand{\ee}{\end{equation}}
\newcommand{\ba}{\begin{eqnarray}}
\newcommand{\ea}{\end{eqnarray}}

\begin{document}

\title{Finite Size Effects on the Chiral Phase Transition of Quantum Chromodynamics}

\author{Shen-Song Wan}
\affiliation{School of Nuclear Science and Technology, Lanzhou University, 222 South Tianshui Road, Lanzhou 730000, China}
\author{Daize Li}
\affiliation{School of Nuclear Science and Technology, Lanzhou University, 222 South Tianshui Road, Lanzhou 730000, China}
\author{Bonan Zhang}
\affiliation{School of Nuclear Science and Technology, Lanzhou University, 222 South Tianshui Road, Lanzhou 730000, China}
\author{Marco Ruggieri}\email{ruggieri@lzu.edu.cn}
\affiliation{School of Nuclear Science and Technology, Lanzhou University, 222 South Tianshui Road, Lanzhou 730000, China}


\begin{abstract}
We study the effect of periodic boundary conditions on chiral symmetry breaking and its restoration in Quantum Chromodynamics.
As an effective model of the effective potential for the quark condensate, we use the quark-meson model,
while the theory is quantized in a cubic box of size $L$.
After specifying a renormalization prescription for the vacuum quark loop,
we study the condensate at finite temperature, $T$, and quark chemical potential, $\mu$.
We find that lowering $L$   
leads to a catalysis of chiral symmetry breaking.
The excitation of the zero mode leads to a jump in the condensate at low temperature and high density, 
that we suggest to interpret as a gas-liquid phase transition that takes place between the chiral symmetry broken phase
(hadron gas) and chiral symmetry restored phase (quark matter).
We characterize  this intermediate phase in terms of the 
increase of the  baryon density, and of the
correlation length of the fluctuations
of the order parameter: for small enough $L$ the correlation domains occupy a substantial
portion of the volume of the system, and the fluctuations are comparable to those in the critical region. For these reasons,
we dub this phase as the {\it subcritical liquid}.
The qualitative picture that we draw 
is in agreement with previous studies based on similar effective models. We also clarify the discrepancy
on the behavior of the critical temperature versus $L$ found in different models.
\end{abstract}



\maketitle

\section{Introduction}

The study of  Quantum Chromodynamics (QCD) at high temperature
and/or baryon density is certainly one of the most active and interesting research topic of
modern high energy physics. It has been predicted by means of  first principles calculations 
that a smooth crossover exists from the low temperature hadron gas phase to a quark-gluon plasma,
at a (pseudo)-critical temperature $T_c\approx 150$ 
MeV \cite{Borsanyi:2010bp,Borsanyi:2010cj,Cheng:2009zi,Bazavov:2011nk,Borsanyi:2013bia}.
This crossover is accompanied by the approximate restoration of chiral symmetry.  
While many studies focused on the QCD phase structure in the infinite volume limit,
it is of a certain interest to investigate the effects of finite size on the critical lines of QCD
in the temperature-baryon chemical potential plane.
Chiral symmetry restoration at finite temperature and density, for quantization 
in a cubic box of size $L$ with periodic or antiperiodic
boundary conditions as well as with standing waves conditions, has been treated in a number of studies,
see \cite{Xu:2019gia,Xu:2019kzy,Almasi:2016zqf,Palhares:2009tf,Braun:2011iz,Braun:2005fj,Kiriyama:2006uh,Braun:2005gy,Wang:2018ovx,
Abreu:2019czp,Magdy:2019frj,Aoki:2002uc,Guagnelli:2004ww,Orth:2005kq,Deb:2020qmx,Zhao:2019ruc,Wang:2018qyq,
Ebert:2010eq,Bhattacharyya:2015zka,Bhattacharyya:2014uxa,Bhattacharyya:2012rp}, 
see \cite{Klein:2017shl} for a review.

We perform a study of chiral symmetry breaking in two-flavor QCD using the renormalized quark-meson (QM) 
model \cite{Skokov:2010sf,Frasca:2011zn,Ruggieri:2013cya},
with quantization in a cubic box of size $L$ with periodic boundary conditions. 
Our work differs from previous calculations based on the same model in the way we treat the vacuum quark loop.
In fact, we include the vacuum term in the thermodynamic potential:
since this is a divergent quantity, care should be put in the regularization and then renormalization of this contribution. 
We follow the regularization procedure of \cite{Xu:2019gia} with Pauli-Villars regulators; 
this regularization is suitable for calculations at finite size
because it cancels the ultraviolet divergences in a very transparent manner and leaves a finite quark loop
both in the infinite and the finite $L$ cases. After the regularization has been done,
we perform renormalization requiring
that in the infinite volume limit, the quark loop does not shift the expectation value of the condensate
and of the $\sigma-$meson mass obtained from the classical potential.
Applying these conditions fixes the counterterms in a way that does not depend of $L$.
Therefore, the predictions of the model at finite $L$ are unaffected by the ultraviolet divergence.

The main purpose of this study is the restoration of chiral symmetry at finite quark chemical potential, $\mu$,
and low temperature. When the size is small enough, $L\lesssim 5$ fm, 
an intermediate phase appears between the chiral symmetry broken phase (the hadron gas) 
and chiral symmetry restored phase (normal quark matter). In this intermediate phase,
the condensate experiences a decrease but its lowering is not enough to restore chiral symmetry.
Therefore, the symmetry breaking pattern in this new phase is the same of the hadron gas.
However, this phase differs from the hadron gas phase because it has a nonzero baryon density;
the change of density, together with the unchanged symmetry breaking pattern, is reminiscent of a
gas-to-liquid phase transition.
Moreover, the correlation domains of the order parameters are larger than those found in the hadron gas
and normal quark matter phases. 
For all these reasons, we call this new phase the 
{\em subcritical liquid phase}.

We also compute $T_c$ versus $L$. Our results  agree with previous studies that implemented effective models
with periodic boundary conditions or infrared cutoffs  \cite{Xu:2019gia,Palhares:2009tf,Wang:2018ovx,Magdy:2019frj}, 
namely $T_c$ increases with $1/L$.
This is understood in terms of the zero mode that contributes to the  condensate when periodic
boundary conditions are used,
as well as of the curvature of the thermodynamic potential which is negative for any $L$
in the QM model.

The plan of the article is as follows. In Section II we briefly review the quark-meson model
and describe the renormalization procedure when periodic boundary conditions are implemented. 
In Section III we discuss the chiral phase transition at small $T$ and large $\mu$.
 In Section IV we present a few results for the chiral restoration at finite $T$. 
 Finally, in Section V we draw our conclusions.
We use the natural units system $\hbar=c=k_B=1$ throughout this article.

\section{The quark-meson model with periodic boundary conditions}

\subsection{The lagrangian density}

The QM model is an effective model of QCD with quarks, $\sigma$ and $\pi$ mesons
(in the two-flavor versions that we consider here).
The meson lagrangian density is
\begin{eqnarray}
{\cal L}_\mathrm{mesons} &=& \frac{1}{2}\left(
\partial^\mu\sigma\partial_\mu\sigma + \partial^\mu\bm\pi\cdot\partial_\mu\bm\pi
\right) \nonumber\\
&&- \frac{\lambda}{4}\left(\sigma^2 + \bm\pi^2 - v^2\right)^2 +h\sigma,\label{eq:ls1_aa}
\end{eqnarray}
where $\bm\pi = (\pi_1,\pi_2,\pi_3)$ corresponds to the pion isotriplet field. This lagrangian density is
invariant under $O(4)$ rotations. On the other hand, as long as $v^2 > 0$ the potential
develops an infinite set of degenerate minima. We choose one ground state, namely
\begin{equation}
\langle\bm\pi\rangle=0,~~\langle\sigma\rangle\neq 0.\label{eq:gs_aa}
\end{equation}
The ground state~\eqref{eq:gs_aa}
breaks the $O(4)$ symmetry down to $O(3)$ since the vacuum is invariant only under the rotations of the pion fields.
This is how chiral symmetry is spontaneously broken in this model.
Besides the spontaneous breaking, chiral symmetry is broken softly but explicitly by the term
$h\sigma$ in the lagrangian density; in fact, the pion mass is $m_\pi^2 = h/F_\pi$.
At zero temperature and in the chiral limit  $\langle\sigma\rangle=F_\pi\approx 93$ MeV 
where $F_\pi$ denotes the pion decay constant in the vacuum.

The quark sector of the QM model is described by the lagrangian density
\begin{equation}
{\cal L}_\mathrm{quarks} = \bar\psi\left(
i\partial_\mu\gamma^\mu - g(\sigma +i\gamma_5 \bm\pi\cdot\bm\tau)
\right)\psi,\label{eq:qlg_aaa}
\end{equation}
where $\bm\tau$ are Pauli matrices in the flavor space. In the ground state~\eqref{eq:gs_aa} quarks get a dynamical (that is,
a constituent) mass given by
\begin{equation}
M = g\langle\sigma\rangle.\label{eq:pppAAA}
\end{equation}
 We notice that in Eq.~\eqref{eq:qlg_aaa} there is no explicit mass term for the quarks. As a matter of fact,
in this effective model the explicit breaking of chiral symmetry   is achieved by
  $h\neq 0$ in Eq.~\eqref{eq:ls1_aa}.
 Although in Eq.~\eqref{eq:qlg_aaa}  there is no explicit mass term, quarks get a constituent mass because of the spontaneous
breaking of the $O(4)$ symmetry in the meson sector: this implies that the quark chiral condensate can be nonzero.
The total lagrangian density is given by
\begin{equation}
{\cal L}_\mathrm{QM} = {\cal L}_\mathrm{quarks} + {\cal L}_\mathrm{mesons}.
\end{equation}
In the following, we will use the notation $\sigma$ to denote both the field and its expectation value,
unless from the context it is not clear which of the two we write about.

\subsection{Renormalized thermodynamic potential in the infinite volume limit}

The mean field effective potential of the QM model in the infinite volume is given by
\begin{equation}
\Omega = U + \Omega_{0,\infty} + \Omega_T,\label{eq:ep1aa}
\end{equation}
where
\begin{equation}
U = \frac{\lambda}{4}\left(\sigma^2 + \bm\pi^2 - v^2\right)^2 - h\sigma\label{eq:ls1_aaMMM}
\end{equation}
is the classical potential of the meson fields as it can be read from Eq.~\eqref{eq:ls1_aa}, and
\begin{equation}
\Omega_{0,\infty} = -2N_c N_f\int\frac{d^3p}{(2\pi)^3} E_p  \label{eq:ls1_aaMMMa}
\end{equation}
is the one-loop quark contribution, with
\begin{equation}
E_p = \sqrt{p^2 +M^2},~~~M=g\sigma.
\end{equation}
Finally, $\Omega_T$ corresponds to the finite temperature quarks contribution that we specify later.
For regularization and renormalization we can limit ourselves to consider the zero temperature limit of $\Omega$,
\begin{equation}
\Omega_0 = U + \Omega_{0,\infty}.\label{eq:ep1}
\end{equation}
Equation~\eqref{eq:ep1} represents the effective potential for the $\sigma$ field computed at one-loop
and after renormalization it corresponds to the renormalized condensation energy, namely the difference between
the energy of the state with $\langle\sigma\rangle\neq 0$ and $\langle\sigma\rangle= 0$ at $T=\mu=0$.

It is instructive to present the renormalization of $\Omega_0$ in the infinite volume system firstly.
In order to do this, we have to regularize the momentum integral in Eq.~\eqref{eq:ls1_aaMMMa}.
We have found that for problems with quantized momenta, in which summations replace integrals, 
renormalization based on the Pauli-Villars (PV) method is the most convenient one.
Therefore, also in the infinite volume limit we use PV regularization and renormalization.
This is the same method used in \cite{Xu:2019gia}.

In the PV scheme we replace Eq.~\eqref{eq:ls1_aaMMMa} with
\begin{equation}
\Omega_{0,\infty} = -2N_c N_f\int\frac{d^3p}{(2\pi)^3} 
\sum_{j=0}^3c_j \left(E_p^2 + j \xi^2\right)^{1/2},  \label{eq:ls1_aaMMMaPV}
\end{equation}
where $\{c_j\}$ is a set of PV coefficients and $\xi$  is the renormalization scale. 
The integration is understood cut at the scale $p=\Lambda$.
The coefficient $c_0=1$ by convention;
the additional three coefficients are needed to remove the quartic, quadratic and log-type divergences of $\Omega_{0,\infty}$
that appear in the limit $\Lambda\rightarrow\infty$.
The PV coefficients
will be chosen so the aforementioned divergences cancel and the final expression does not depend on $\Lambda$.
Performing the integration, it is an easy exercise to see that the choice $c_1=-3$, $c_2=3$ and $c_3=-1$ is
enough to cancel all the divergences, in agreement with \cite{Xu:2019gia}. The resulting finite expression
is
\begin{eqnarray}
\Omega_{0,\infty} &=& \frac{3N_c N_f}{16\pi^2}
\xi^4\log\frac{(M^2 + \xi^2)(M^2 + 3\xi^2)^3}{(M^2 +2 \xi^2)^4}\nonumber\\
&+&\frac{6N_c N_f}{16\pi^2}
\xi^2 M^2\log\frac{(M^2 + \xi^2)(M^2 + 3\xi^2)}{(M^2 +2 \xi^2)^2}\nonumber\\
&+&\frac{N_c N_f}{16\pi^2}
M^4\log\frac{(M^2 + \xi^2)^3(M^2 + 3\xi^2)}{M^2(M^2 +2 \xi^2)^3}.\nonumber\\
&&\label{eq:desperate12}
\end{eqnarray}

Although Eq.~\eqref{eq:desperate12} is finite, it potentially can shift the location of the minimum of the classical potential
as well as the mass of the $\sigma-$meson in the vacuum, $m_\sigma$.
While this would be not a problem since it would require a mere change of the parameters $\lambda$ and $v$,
it is easier to work assuming that the quark loop does not shift these quantities.
To this end, we add two counterterms,
\begin{equation}
\Omega_\mathrm{c.t.} = \frac{\delta v}{2}M^2 + \frac{\delta\lambda}{4}M^4,\label{eq:ctUUU}
\end{equation}
and we impose the renormalization conditions
\begin{eqnarray}
&&\left.\frac{\partial (\Omega_{0,\infty} + \Omega_\mathrm{c.t.})}{\partial M}\right|_{M=gF_\pi}=0,\label{eq:sh4aF}\\
&&\left.\frac{\partial^2 (\Omega_{0,\infty} + \Omega_\mathrm{c.t.})}{\partial M^2}\right|_{M=gF_\pi}=0.\label{eq:sh5aF}
\end{eqnarray}
The first condition imposes  that the quark loop does not change the
location of the minimum of the classical potential, $\sigma=F_\pi$,
while the second states that the loop does not shift $m_\sigma$.
The coefficients of the two counterterms can be computed easily,
\begin{eqnarray}
\delta v &=& -\frac{3N_c N_f}{4\pi^2}\xi^2\log\frac{(g^2F_\pi^2 + \xi^2)(g^2F_\pi^2 + 3\xi^2)}{(g^2F_\pi^2 + 2\xi^2)^2},
\label{eq:ctdv1}\\
\delta\lambda &=&-\frac{N_c N_f}{4\pi^2}
\log\frac{(g^2F_\pi^2 + \xi^2)^3(g^2F_\pi^2 + 3\xi^2)}{g^2F_\pi^2(F_\pi^2 + 2\xi^2)^3}.\label{eq:ctdl1}
\end{eqnarray}
The renormalized quark loop is thus given by
\begin{equation}
\Omega_{0,\infty}^\mathrm{ren} = \Omega_{0,\infty} + \Omega_\mathrm{c.t.}.
\label{eq:rty}
\end{equation}

\subsection{Renormalized thermodynamic potential in the finite volume case}
In a finite volume $V=L^3$ the $i-$component of momentum is quantized according to
\begin{equation}
p_i = \frac{2\pi}{L}n_i,~~~n_i=0,\pm1,\pm2,\dots; \label{eq:mom_quant}
\end{equation}
this leads to the obvious replacements
\begin{eqnarray}
\int \frac{d^3p}{(2\pi)^3} &\rightarrow &\frac{1}{V}\sum_{n_x,n_y,n_z,}, \label{eq:mom_repl}\\
E_p &\rightarrow &E_n = \left(
M^2 + \frac{4\pi^2}{L^2}n
\right)^{1/2},\label{eq:Ep_repl}
\end{eqnarray}
with $n=n_x^2 + n_y^2 + n_z^2$.
Instead of $\Omega_{0,\infty}$ we have  
\begin{eqnarray}
\Omega_{0,L}(\Lambda) &=&-\frac{2N_c N_f}{L^3}|M|
\nonumber\\
&& -\frac{2N_c N_f}{L^3}\sum_{n=1}^{a}r_3(n)\sqrt{\frac{4\pi^2}{L^2}n+M^2},
\label{eq:alpha_2}
\end{eqnarray}
where the subscript $L$ reminds that the potential is computed assuming quantization in a box with volume $L^3$;
the first addendum on the right hand side of Eq.~\eqref{eq:alpha_2} is the zero mode contribution.
We have put $a=\Lambda^2 L^2/4\pi^2$ 
where $\Lambda$ is an UV cutoff that will disappear after 
PV renormalization; 
$r_3(n)$ denotes the sum-of-three-squares function, 
that counts how many ways it is possible to form $n$ as the sum of the squares of three integers: 
it corresponds to the degeneracy of the level with a given $n$.
Similarly, instead of Eq.~\eqref{eq:ep1} we have
\begin{equation}
\Omega_0(\Lambda) = U + \Omega_{0,L}(\Lambda).\label{eq:ep1_L}
\end{equation}
This is the bare potential that is divergent and needs renormalization:
to make this evident we have made the dependence of $\Lambda$ explicit.

The  renormalization  of the thermodynamic potential is performed
following the PV method delineated in the infinite volume case.
Firstly, we introduce a set of PV coefficients and 
replace Eq.~\eqref{eq:alpha_2} with
\begin{eqnarray}
\Omega_{0,L} &=& -\frac{2N_c N_f}{L^3}|M|\nonumber\\
&&-\frac{2N_c N_f}{L^3}\sum_{n=1}^{a}r_3(n)
\sum_{j=0}^3 c_j\sqrt{E_n^2 + j \xi^2}.
\label{eq:alpha_2a}
\end{eqnarray}
Using the coefficients determined in the infinite volume limit is enough to get a finite expression in the $\Lambda\rightarrow\infty$
limit. This can be proved by brute force numerically, but can also be understood analytically as follows.
For studying the UV divergence, we put  $X_j^2 = M^2 + j\xi^2$ 
and we extract the $O(X_j^2)$ and $O(X_j^4)$ terms from  $\Omega_{0,L}$,
that will bring a quadratic and log-type divergence respectively. Thus we can write
\begin{eqnarray}
\Omega_{0,L} &=& -\frac{2N_c N_f}{L^3}|M|\nonumber\\
&&-\frac{2N_c N_f}{L^3}\sum_{n=1}^{a}r_3(n)
\sum_{j=0}^3 c_j\left[a_2 X_j^2 + a_4 X_j^4\right]\nonumber\\
&&+~\mathrm{UV~finite~terms},
\label{eq:alpha_2a2}
\end{eqnarray}
where 
\begin{eqnarray}
a_{2} &=&\frac{L}{4\pi n^{1/2}},\\
a_{4} &=& -\frac{L^3}{64\pi^3 n^{3/2}}.
\end{eqnarray}
By virtue of a numerical calculation we prove that in the large $a$ limit
\begin{eqnarray}
\sum_{n=1}^a \frac{r_3(n)}{4\pi n^{1/2}} &\approx &\frac{a}{2},\\
\sum_{n=1}^a \frac{r_3(n)}{4\pi n^{3/2}} &\approx & \frac{\log a}{2},
\end{eqnarray}
which allow to write
\begin{eqnarray}
\Omega_{0,L} &=& -\frac{2N_c N_f}{L^3}|M|\nonumber\\
&&-\frac{2N_c N_f}{L^3} 
\sum_{j=0}^3 c_j\left[a_2 X_j^2\frac{L a}{2} - a_4 X_j^4\frac{L^3 \log a}{32\pi^2}\right]\nonumber\\
&&+~\mathrm{UV~finite~terms}.
\label{eq:alpha_2a3}
\end{eqnarray}
The above equation shows that $a_2$ and $a_4$ multiply the quadratic and log-type divergence respectively. 
Using the PV coefficients it is easy to prove that 
\begin{eqnarray}
\sum_{j=0}^3 c_j X_j^2 &=& 0,\\
\sum_{j=0}^3 c_j X_j^4 &=& 0,
\end{eqnarray}
while the $O(X_j^6)$ term is nonzero and UV-finite. 
Therefore, the PV regulator cancels the UV divergence of $\Omega_{0,L}$
leaving a UV-finite, $\xi-$dependent term.

The counterterms that we have fixed in the infinite volume case can be used here as well:
they will implement the conditions that in the large volume limit, we recover
$\sigma=F_\pi$ and the $m_\sigma$ fixed by the classical potential.
On the other hand, for a finite $L$ the $\Omega_{0,L}$ can shift both the location
of the minimum of the total potential and $m_\sigma$. Therefore,
for finite $L$ we will use
\begin{equation}
\Omega_{0,L}^\mathrm{ren} = \Omega_{0,L} + \Omega_\mathrm{c.t.},
\label{eq:rtyAA}
\end{equation}
with $\Omega_\mathrm{c.t.}$ specified bt Eq.~\eqref{eq:ctUUU} with counterterms
given by Eqs.~\eqref{eq:ctdv1} and~\eqref{eq:ctdl1}. 
Taking into account the classical potential, the renormalized thermodynamic potential
in the vacuum at finite $L$ is
\begin{equation}
\Omega^\mathrm{ren} = U + \Omega_{0,L}+\Omega_\mathrm{c.t.}.
\label{eq:renpot_1}
\end{equation}

We close this subsection with a short comment on the choice of $\xi$. 
In principle, we could change this arbitrarily  at a given $L$
by requiring that the total derivative of $\Omega$ with respect to $\xi$, $d\Omega/d\xi$,  is zero. 
This would amount to solve a 
Renormalization Group-like equation in which the $\partial\Omega/\partial\xi$ is balanced by terms
proportional to $\partial\lambda/\partial\xi$ and $\partial g/\partial\xi$ so that $d\Omega/d\xi=0$. 
Solving this equation is well beyond the purpose of the study we want to do here.
Therefore, for a given $L$ we have limited ourselves to inspect the ranges of $\xi$ that do not change $\Omega$ 
too much. For large $L$ we have found that $\xi$ can be arbitrarily large. 
On the other hand, for small $L$ we have found that $\Omega$ is quite insensitive to the specific value of $\xi$  
as long as $\xi \lesssim \gamma F_\pi $ with $\gamma=O(1)$. Therefore, we  fix $\xi=F_\pi$ in this work.

\subsection{The total thermodynamic potential}
The finite temperature thermodynamic potential does not need any particular treatment:
in infinite volume it is given by the standard relativistic fermion gas contribution, namely
\begin{equation}
\Omega_T = -2N_c N_f T\sum_{s=\pm 1} \int\frac{d^3p}{(2\pi)^3}\log\left(1 + e^{-\beta(E_p-s\mu)}\right),
\end{equation}
where $\mu$ corresponds to the chemical potential. 
In finite volume we replace the above equation with
\begin{equation}
\Omega_T = -2N_c N_f \frac{T}{L^3}\sum_{s=\pm 1}\sum_{n}r_3(n)
\log\left(1 + e^{-\beta(E_n-s\mu)}\right).\label{eq:iow}
\end{equation}
Putting all together, we get the renormalized thermodynamic potential of the QM model in a volume $V=L^3$
with periodic boundary conditions, that is
\begin{equation}
\Omega = U + \Omega_{0,L}+\Omega_\mathrm{c.t.}+ \Omega_T.
 \label{eq:tot_om_fs}
\end{equation}
For each value of $\mu$ and $T$ we determine $\sigma$ by looking for the global minimum of $\Omega$. 

\subsection{Parameters of the classical potential}
The 
renormalization procedure outlined above has the advantage that does not require a shift of the 
parameters of the classical potential both in the infinite volume and in the finite size cases. 
Therefore, these parameters can be computed from $U$ and are not affected by $L$.
These can be computed easily from the conditions that $\partial U/\partial\sigma=0$ and
$\partial^2 U/\partial\sigma^2 = m_\sigma^2$, where the derivatives are understood 
computed at $\sigma=F_\pi$.
Limiting ourselves to write concrete expressions in the limit $h\rightarrow 0$ we find
\begin{eqnarray}
v &=& F_\pi  - \frac{h}{m_\sigma^2},\\
\lambda  &=& \frac{m_\sigma^2 }{2 F_\pi^2 } - \frac{h}{2F_\pi^3 }.
\end{eqnarray}


\section{Chiral phase transition at small temperature\label{Sec:llp}}

In this section we present the results for the 
condensate and phase structure at low temperature
and high density, which is the domain in which the most interesting finite size effects appear. 
Our parameters set is $m_\sigma=700$ MeV, $F_\pi=93$ MeV,
$m_\pi=138$ MeV, $h=F_\pi m_\pi^2$.
Finally,  we take $\xi=F_\pi$ and we assume $M=g F_\pi=335$ MeV at $\mu=T=0$ which gives $g=3.6$.

\subsection{Condensate, $m_\sigma$ and $m_\pi$ versus size in the vacuum}

\begin{figure}[t!]
\begin{center}
\includegraphics[width=0.45\textwidth]{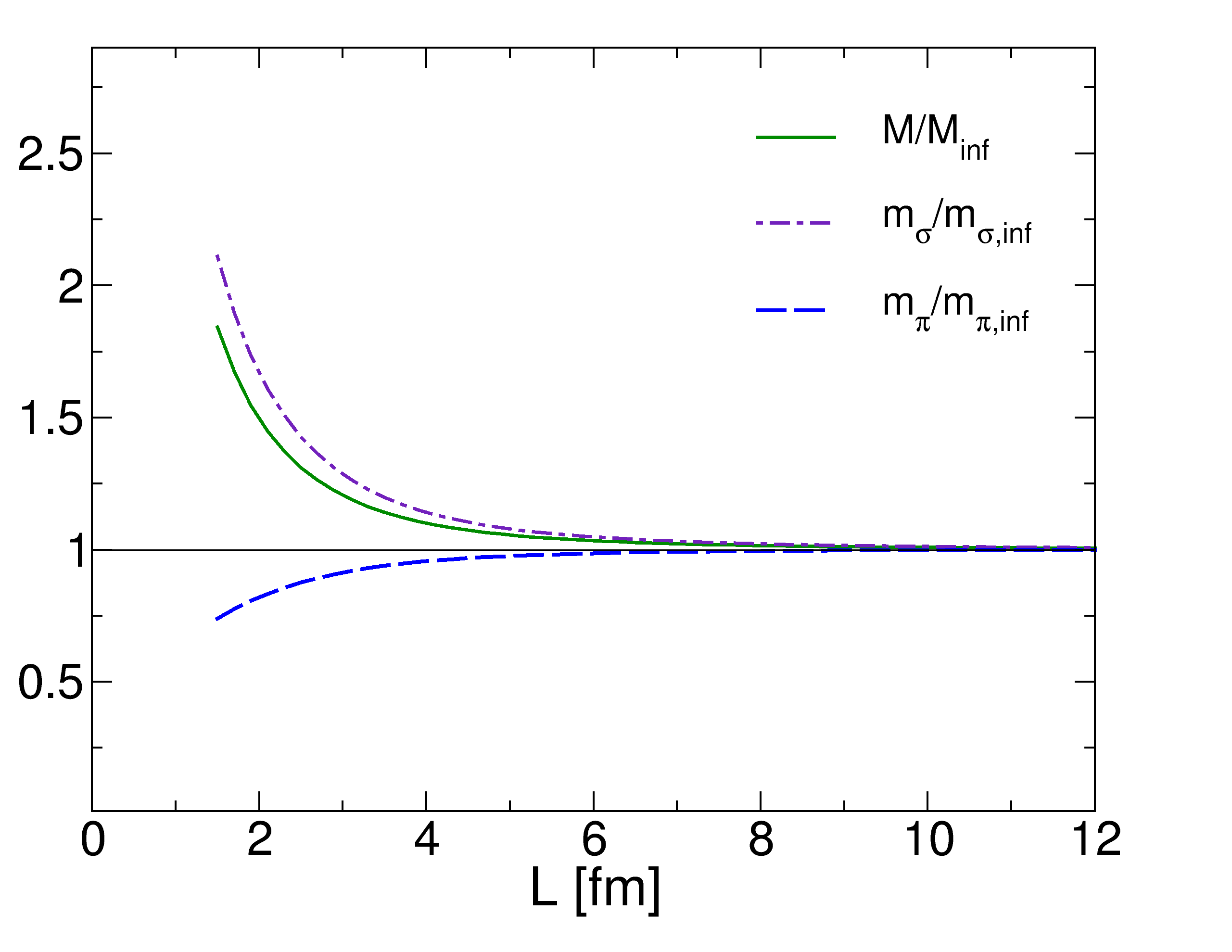}
\end{center}
\caption{\label{Fig:vacvac}Quark mass 
and $m_\sigma$ versus $L$ in the vacuum. 
Both quantities are measured in units of the infinite volume cases.
Renormalization scale is $\xi=F_\pi$.}
\end{figure}

To begin with, we present the behavior of the condensate, of the $\sigma-$meson mass and pions mass
versus $L$ in the vacuum: the results are summarized in Fig.~\ref{Fig:vacvac}.
All quantities are measured in units
of the infinite volume cases.  We find that both $M$ and $m_\sigma$ increase with lowering $L$,
while $m_\pi$ decreases.
The effects of a finite $L$ on the physical quantities become noticeable for $L\lesssim 3$ fm.
Qualitatively the results of the renormalized QM model agree with   
the NJL model calculations \cite{Xu:2019gia}.

\subsection{Transitions at small temperature}

\begin{figure}[t!]
\begin{center}
\includegraphics[width=0.45\textwidth]{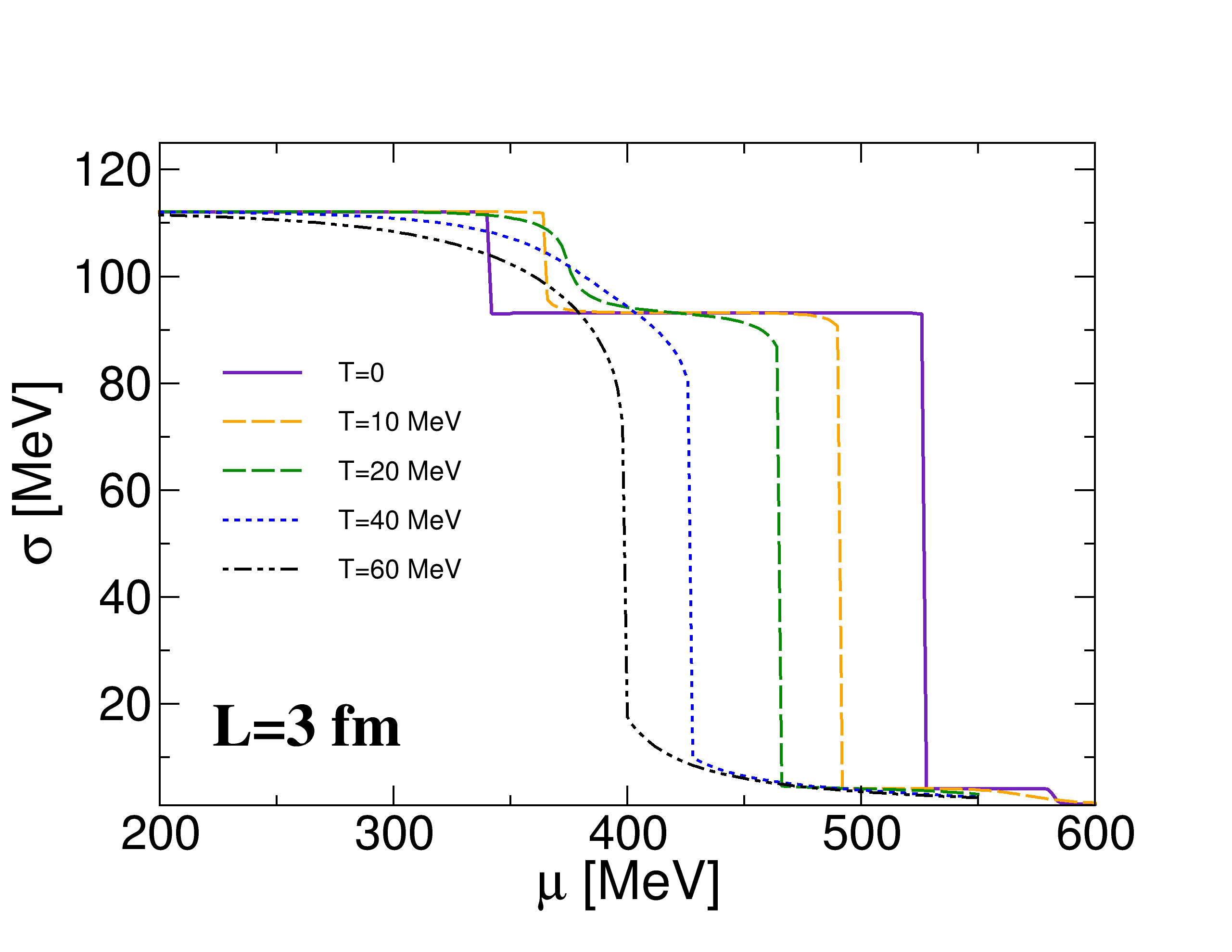}\\
\includegraphics[width=0.45\textwidth]{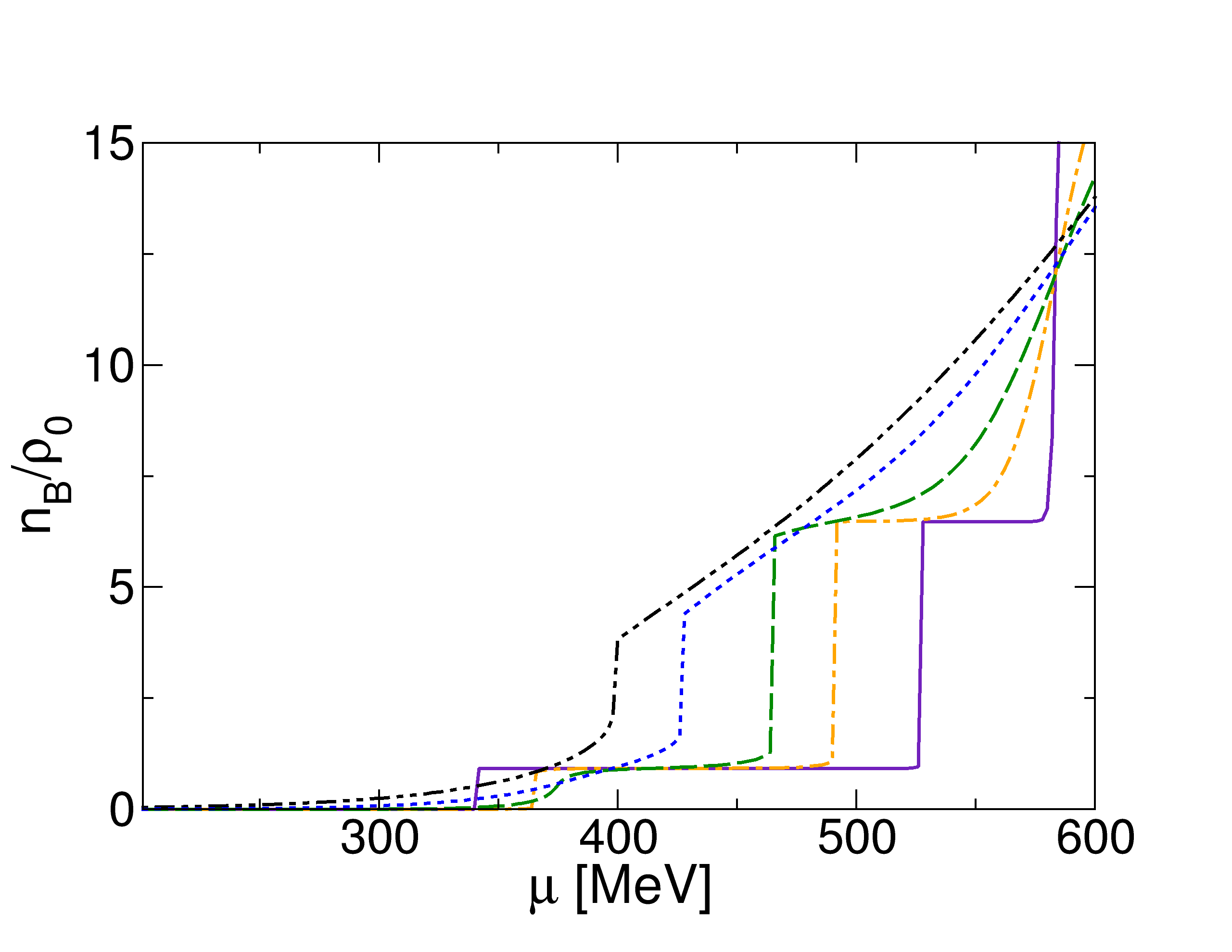}
\end{center}
\caption{\label{Fig_cc} Condensate (upper panel)
and $n_B/\rho_0$ (lower panel) versus chemical potential, for several values of $T$.
Calculations correspond to $L=3$ fm.}
\end{figure}

In the upper panel of Fig.~\ref{Fig_cc} we plot the condensate versus $\mu$ for several temperatures
and for $L=3$ fm. The condensate has been computed by a global minimization procedure of $\Omega$
for any $(\mu,T)$.
For $T=0$  $\sigma$ has a discontinuity for $\mu\equiv\mu_1\approx 340$ MeV and drops down 
from its value in the vacuum 
to a smaller, still substantial value $\sigma_1\approx 93$ MeV.
This discontinuity agrees with the one found in \cite{Xu:2019gia} where it has been shown to be driven by the zero mode
(a direct calculation within the QM model confirms this).
Despite the discontinuity of $\sigma$, chiral symmetry is still spontaneously broken 
for $\mu>\mu_1$ since the value of the condensate
is still large. 
Increasing $\mu$ up to a second critical value $\mu\equiv\mu_2\approx 526$ MeV there is another
jump of $\sigma$ to a $\sigma_2\approx 5$ MeV:
it is fair to identify this discontinuity with the restoration of chiral symmetry. 
It is easy to verify that the phase transition happens when the chemical potential is large enough
to populate the first excited state, $n=1$: in fact,
using $\sigma =\sigma_2$  the energy of this state is
$E_1 = \sqrt{g^2\sigma^2 + 4\pi^2/L^2}\approx 415$ MeV, therefore the state $n=1$ can be populated
for $\mu \gtrsim \mu_2$.
Increasing the temperature, the chiral phase transition and the jump of the condensate approach each other;
moreover, the discontinuity of $\sigma$ at $\mu=\mu_1$ is smoothed by temperature becoming a crossover.

At low enough $L$ and low $T$  an intermediate phase appears between the chiral symmetry breaking phase
at low $\mu$, namely the hadron gas, and the high density phase that is quark matter in which chiral symmetry
is restored. Even though the passage from the chiral symmetry broken phase to the intermediate one is not 
a phase transition because symmetries are broken in the same way in the two phases, for the sake of simplicity we   
adopt the term transition to discuss also the change of the condensate for $\mu=\mu_1$.
In fact, we aim to interpret this transition as a gas-to-liquid phase transition.

It is interesting to examine the behavior of number density around the two transitions.
To this end we define the baryon density, $n_B$, as
\begin{equation}
n_B = \frac{n_u + n_d}{3},\label{eq:bnm3} 
\end{equation}
where $n_{u}$ and $n_d$ denote the densities of $u$ and $d$ quarks respectively; it is straightforward to prove that 
\begin{equation}
n_B =  -\frac{1}{3}\frac{\partial\Omega}{\partial\mu}.\label{eq:bnm_inter}
\end{equation}
In the lower panel of Fig.~\ref{Fig_cc} we plot $n_B$ versus $\mu$ for several values of $T$ and $L=3$ fm;
baryon density is measured in units of the nuclear saturation density, $\rho_0= 0.16$ fm$^{-3}$.
At small temperature, $n_B$ experiences a first jump from zero to $n_B\approx 0.95\rho_0\equiv n_B^{(1)}$
for $\mu=\mu_1$, stays constant then experiences another jump to $n_B\approx 6.46\rho_0\equiv n_B^{(2)}$
for $\mu=\mu_2$.
 
The dependence of $n_B$ on $\mu$ at small temperature can be easily understood.
As a matter of fact, at zero (as well as very small but finite) temperature,
if $\mu$ is large enough to excite the zero as well as the first mode,
 we have
\begin{equation}
n_B \approx \frac{2N_c N_f}{3V}
\left[
\theta(\mu-M) + 6\theta(\mu-\sqrt{M^2 + 4\pi^2/L^2})
\right];\label{eq:bnm5}
\end{equation}
the first addendum in the right hand side of the above equation is the contribution of the zero mode,  
while the second addendum
corresponds to the first excited state 
counted with its degeneracy $r_3(1)=6$. 
Baryon density is constant for 
$M <\mu < \sqrt{M^2 + 4\pi^2/L^2}$ where only the zero mode contributes;
analogously, density is constant also for larger values of $\mu$ until the second excited state can be populated.
This is qualitatively different from the behavior $n_B \propto (\mu-M)^{3/2}$ of a 
relativistic ideal massive gas, because for a finite size system there are only discrete modes in the spectrum,
and if temperature is low enough only few of them can be occupied giving rise to $\Omega_T\propto\mu$.
Only when the degeneracy becomes large  one can approach the continuum limit and eventually recover the aforementioned
dependence of the density on the chemical potential.

We suggest a similitude between the jump of the chiral condensate and a liquid-gas phase transition.
As a matter of fact, chiral symmetry is not restored at $\mu_1$,
therefore the pattern of symmetry breaking is the same at low and intermediate $\mu$.
In addition to this,
the quark number density at the first jump has a net increase, 
and the transition is sharp at low temperature then becomes smooth at higher temperatures.
These aspects characterize a liquid-gas phase transition,
which corresponds to a change in density and not to a change in the pattern of spontaneous symmetry breaking.

\subsection{Correlation length of the fluctuations of the order parameter}

We can further characterize this liquid-gas-like jump of the condensate at $\mu_1$ by means of the correlation length
of the static fluctuations of the condensate, that are carried by the $\sigma-$meson. 
To do this, firstly we have to compute the in-medium masse of the $\sigma-$meson.
Computing this within the quark-meson model is a well established procedure, see for example \cite{Castorina:2020vbh}
and references therein, and is straightforward when two-loop contributions are neglected and 
the Hatree approximation is used to compute the effective 2-particle-irreducible potential. Within these approximations we have
$M_\sigma^2 = \partial^2\Omega/\partial\sigma^2$,
where the second derivative is understood at the global minimum of $\Omega$.
A similar equation holds for the $\pi-$mesons, $M_\pi^2 = \partial^2\Omega/\partial\pi^2$,
where $\Omega$ can be augmented with the pion field by the obvious replacement $\sigma^2\rightarrow\sigma^2 + \pi^2$
in all but the $h\sigma$ terms and in the zero mode contribution $M\rightarrow \sqrt{M^2 + g^2\pi^2}$.

\begin{figure}[t!]
\begin{center}
\includegraphics[width=0.45\textwidth]{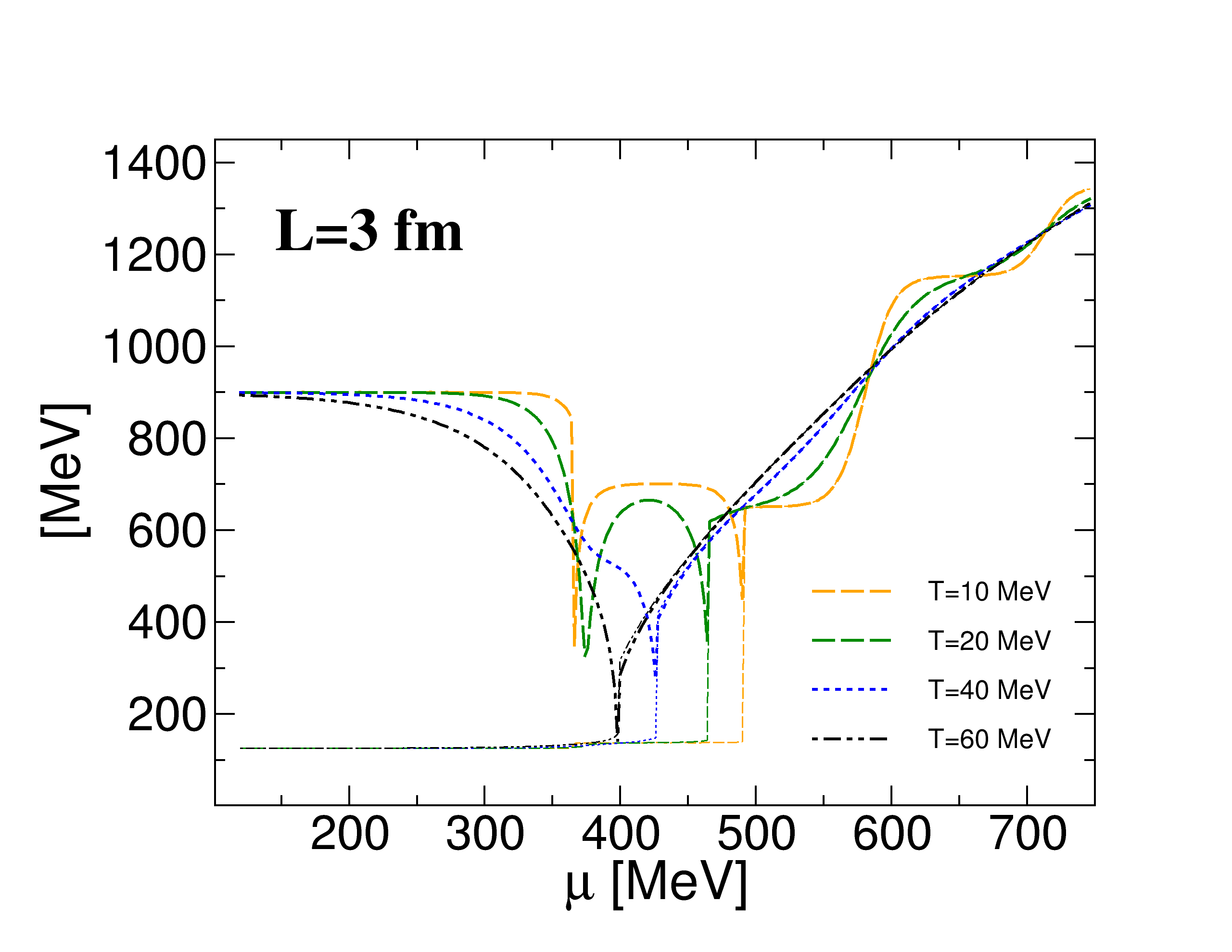}
\end{center}
\caption{\label{Fig:3fm}In-medium $\pi-$meson and $\sigma-$meson masses
versus $\mu$, for several temperatures and $L=3$ fm. Thin lines correspond to $M_\pi$ while thick lines
denote $M_\sigma$.
}
\end{figure}

In   Fig.~\ref{Fig:3fm} we plot $M_\sigma$ and $M_\pi$ versus $\mu$,
for several values of $T$. At $\mu=\mu_1$ $M_\sigma$ drops down. However, 
$M_\pi$ is almost insensitive to the jump of the condensate. This confirms that
the first jump at $\mu_1$ should not be identified with a real phase transition.
At $\mu=\mu_2$ where the condensate drops down to almost zero,
$M_\sigma$ and $M_\pi$ join and increase, signaling that the $O(4)$ symmetry is restored and these particles
become heavy enough that decouple from the low energy spectrum dominated by the quarks. 
This confirms that it is $\mu_2$ that has to be identified with chiral symmetry restoration.

\begin{figure}[t!]
\begin{center}
\includegraphics[width=0.45\textwidth]{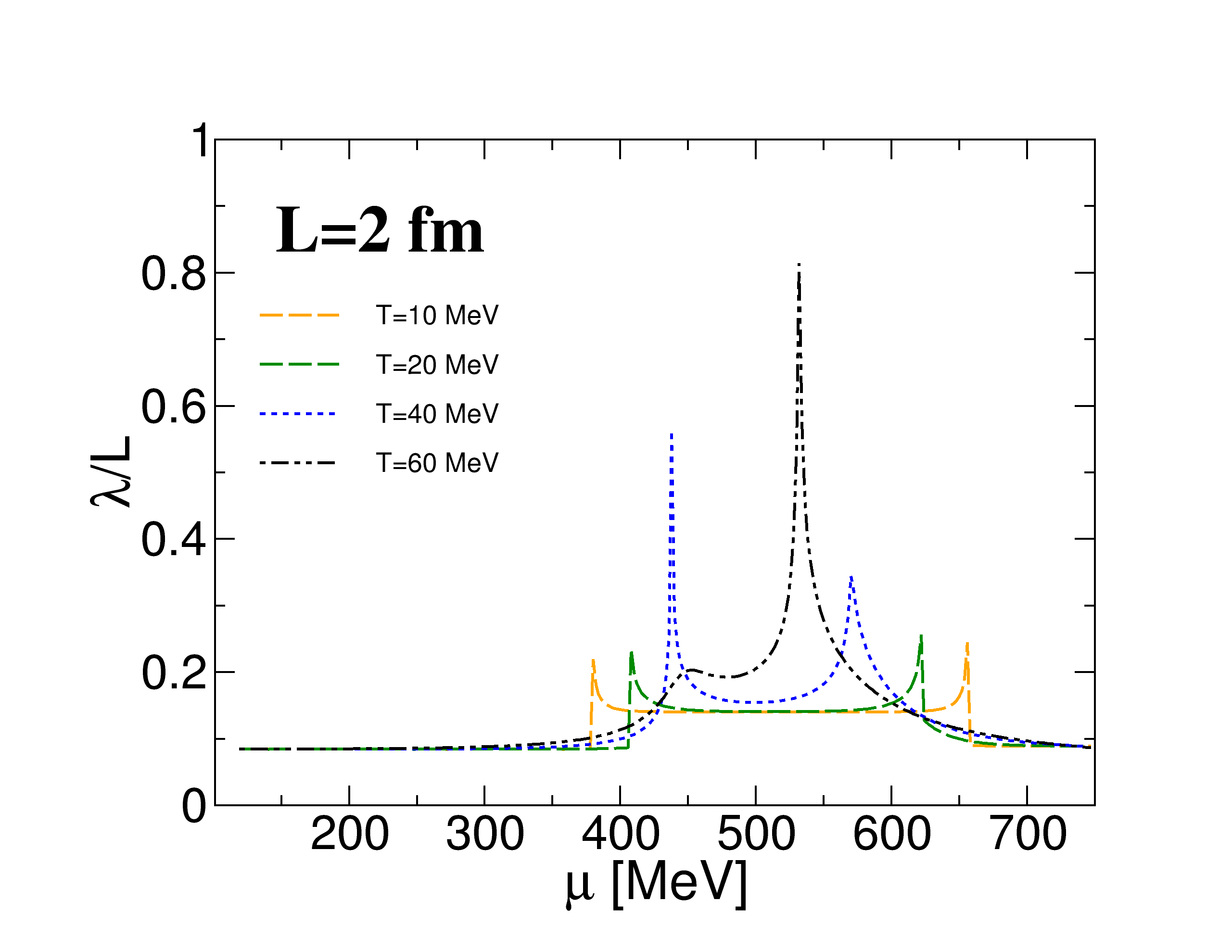}\\
\includegraphics[width=0.45\textwidth]{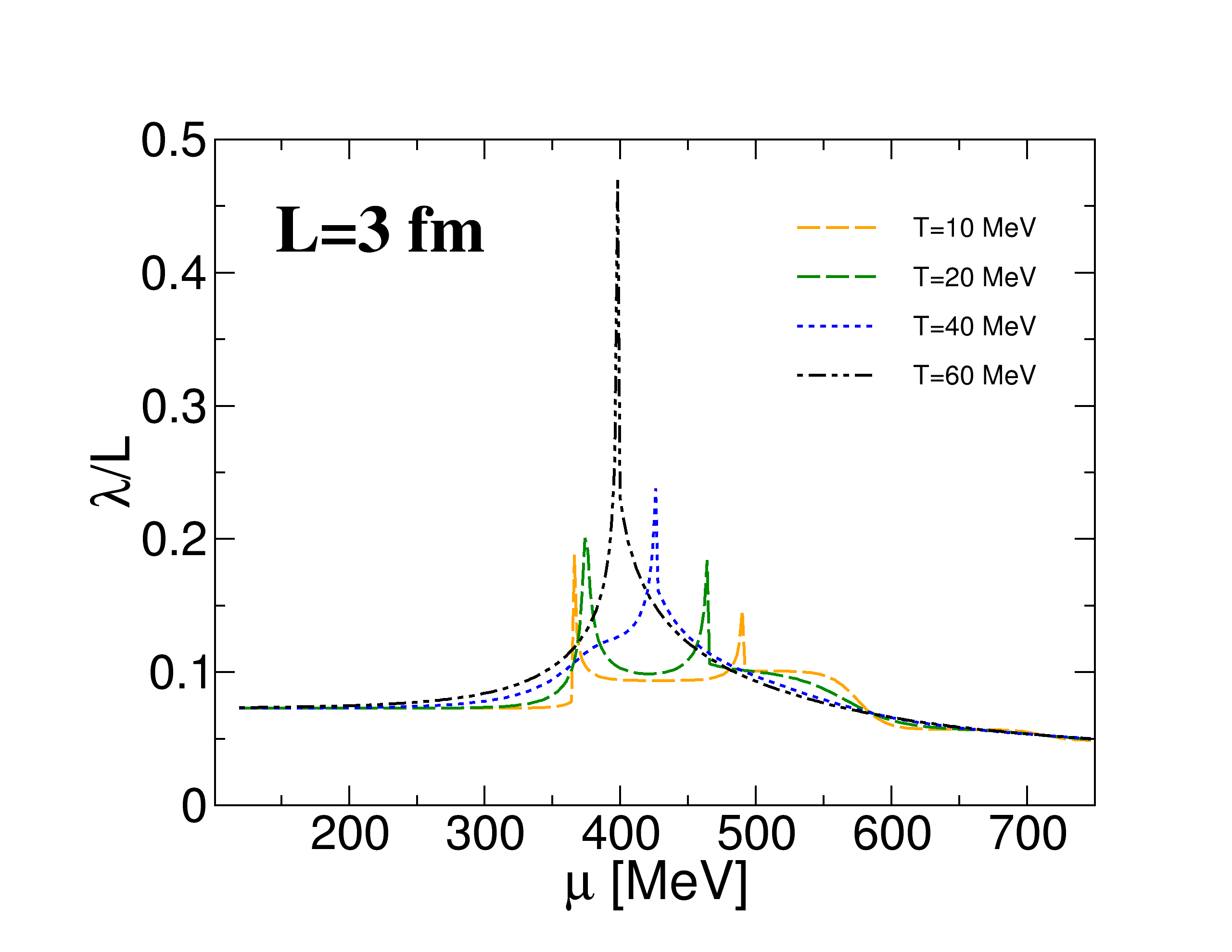}\\
\includegraphics[width=0.45\textwidth]{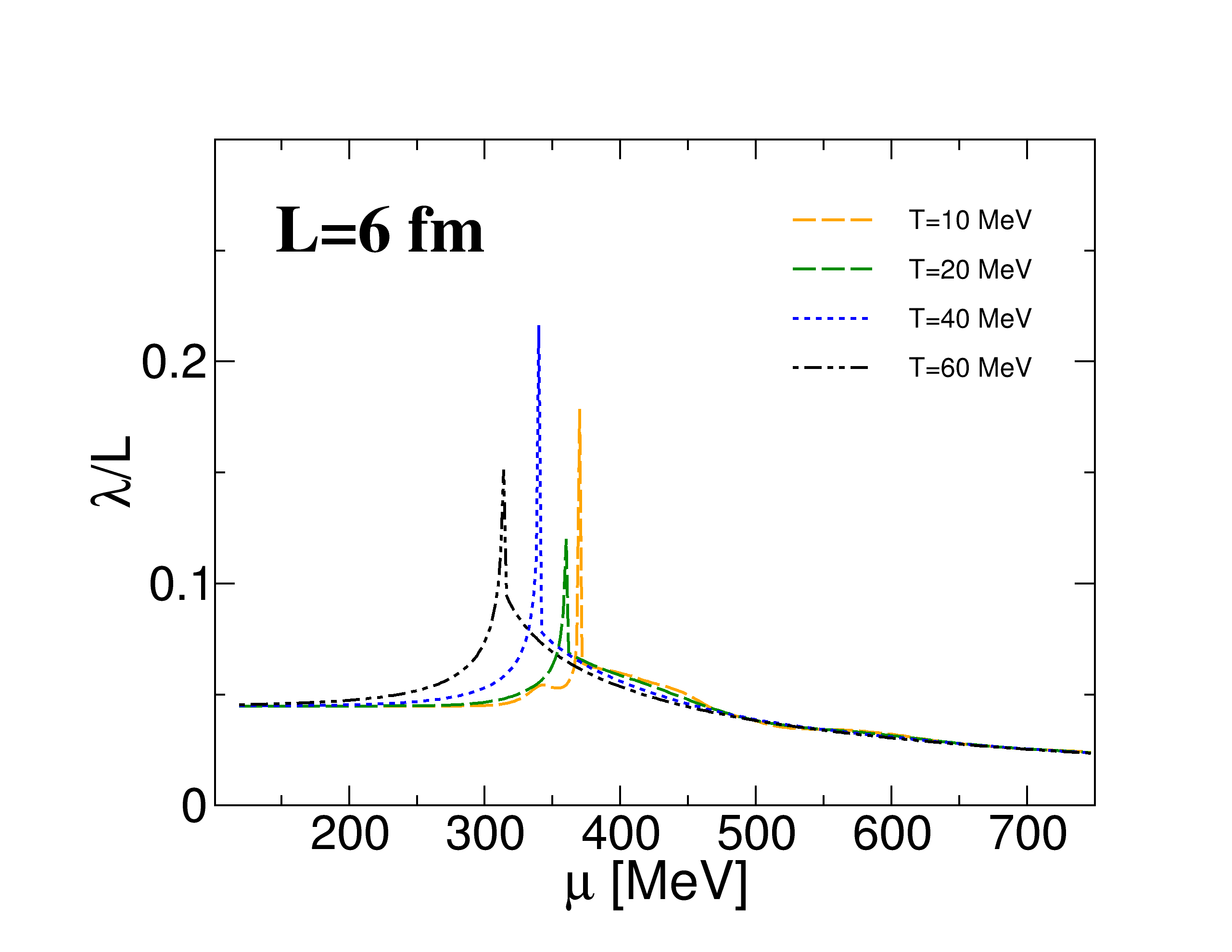}
\end{center}
\caption{\label{Fig:3fmCOR}Correlation length of the fluctuations of the order parameter versus $\mu$, for several temperatures.
}
\end{figure}

We can compute the correlation length of the time independent fluctuations of the condensate, 
$\lambda$, that are transported by the $\sigma-$meson. In fact, for time independent fluctuations the effective action
of the $\sigma-$meson is formally equivalent to that of a Ginzburg-Landau theory
for a scalar order parameter with positive squared mass, and it is textbook matter that in this case the correlation length
of fluctuations is nothing but  $\lambda=1/M_\sigma$. We show $\lambda/L$ versus $\mu$ in  
Fig.~\ref{Fig:3fmCOR} for three representative values of $L$.
For $L=2$ fm, at $T=10$ MeV and $\mu=\mu_1$ the correlation length increases and stays constant
up to $\mu=\mu_2$; for larger values of $\mu$ it decreases and approaches the value it has in the hadron gas phase.
The correlation length is frozen in the intermediate phase, due to the fact that only the zero mode is excited. 
Qualitatively, this happens also for $T=20$ MeV and $T=40$ MeV.
Also notice that in this phase $\lambda/L\approx 0.2$ meaning that one correlation volume occupies
about the twenty percent of the volume of the system; around the two transitions $\lambda$
develops two peaks; in particular, for $T=40$ MeV we find that $\lambda/L=O(1)$ that implies
that the system is close to criticality. This is what we would expect at a critical endpoint.
Finally, for $T=60$ MeV there is only one peak of $\lambda$ in agreement with the fact that
the transition to the intermediate phase is smoothed by the temperature;
nevertheless, $\lambda$ experiences a net increase in comparison with the value at small $\mu$,
then again $\lambda/L=O(1)$ at the chiral phase transition.
The qualitative picture is the same at $L=3$ fm, see the middle panel of   Fig.~\ref{Fig:3fmCOR},
while for a larger value of $L$ the double peak structure as well as the intermediate phase disappear,
see the lower panel of Fig.~\ref{Fig:3fmCOR}.

The results  summarized in Fig.~\ref{Fig:3fmCOR} allow to understand better the intermediate phase.
As a matter of fact, we learn that beside the characterization of this phase in terms of the baryon density
discussed in the previous subsection,
we can distinguish it from the vacuum and the high density normal quark matter also looking at
the fluctuations of the order parameter.
In particular, the correlations of the order parameters are substantially larger than those in the vacuum 
and in the quark matter phase at high $\mu$, and do not change by changing $\mu$
in this region:  correlation volumes are frozen due to the fact that only the zero mode
is excited. Moreover, for small $L$ 
the correlation volumes occupy a substantial portion of the total volume of the system.
These facts, together with the increase of density and
the symmetry pattern that is unchanged at $\mu=\mu_1$, 
suggest the name of {\it subcritical liquid} for this intermediate phase.

\subsection{Catalysis of chiral symmetry breaking at low temperature}

\begin{figure}[t!]
\begin{center}
\includegraphics[width=0.45\textwidth]{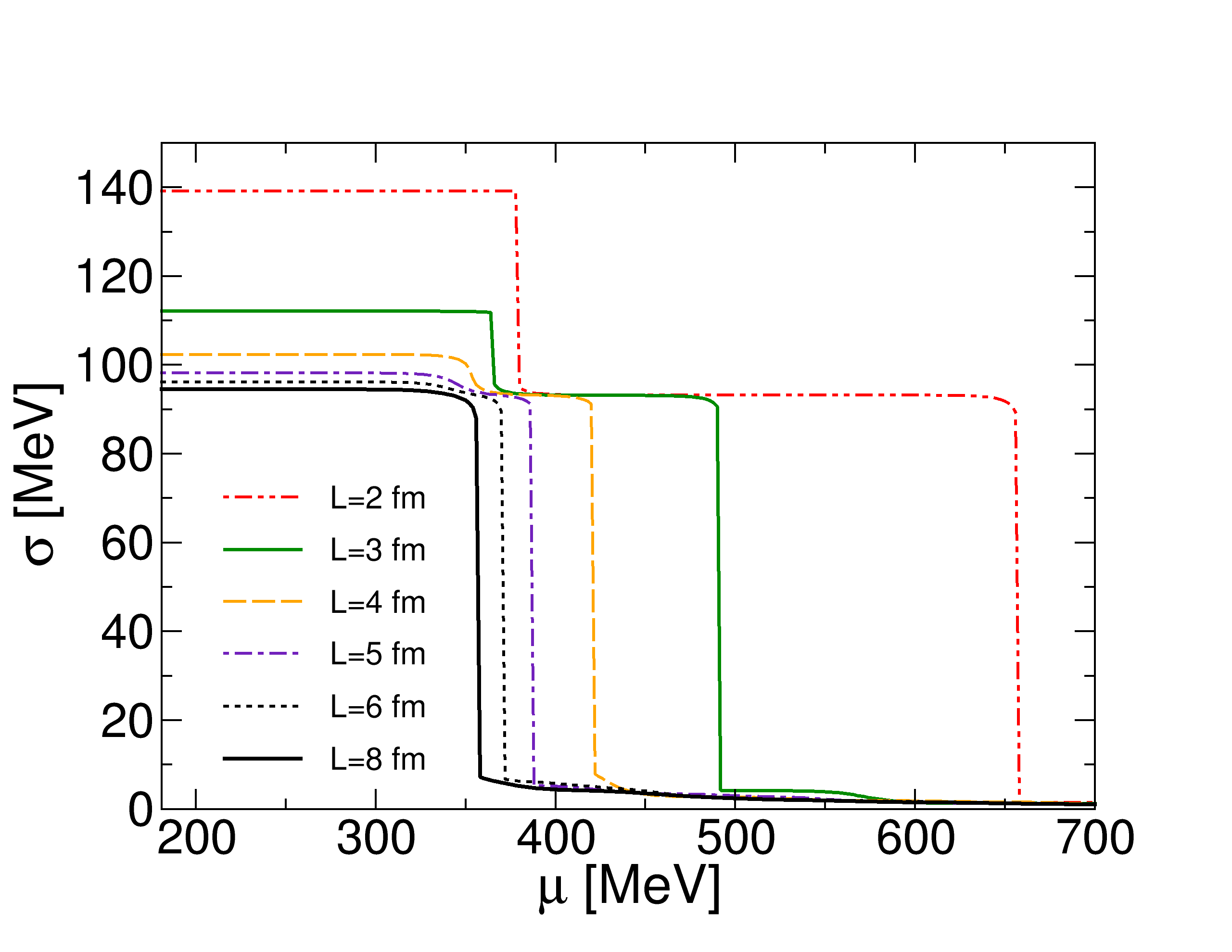}
\end{center}
\caption{\label{Fig:moresize} Condensate versus chemical potential, for several values of $L$
and $T=10$ MeV.}
\end{figure} 

In Fig.~\ref{Fig:moresize} we plot the condensate versus $\mu$ at $T=10$ MeV, for several values of $L$.
Finite size effects are noticeable up to $L\approx 5$ fm although in this case the transition to the subcritical liquid phase
is minor. For $L=6$ fm no sign of the subcritical liquid is found, and comparing the results of $L=6$ fm and $L=8$ fm
we notice that the effect of the finite size is almost gone and a continuum limit is reached.
The results collected in Fig.~\ref{Fig:moresize}  show that lowering the size catalyzes the spontaneous chiral symmetry 
breaking by enlarging the subcritical liquid region.  

\begin{figure}[t!]
\begin{center}
\includegraphics[width=0.45\textwidth]{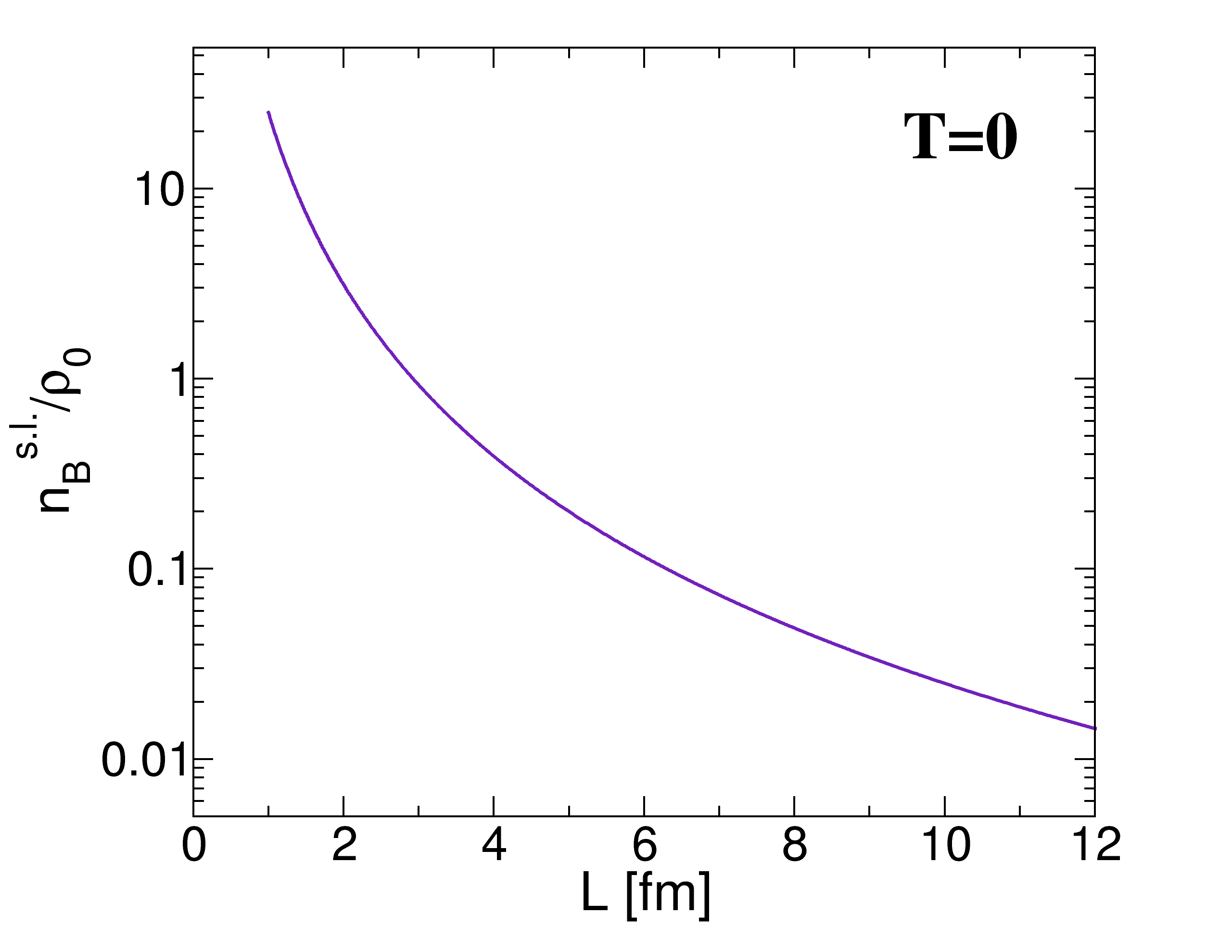}
\end{center}
\caption{\label{Fig:nbsl}$n_B^\mathrm{s.l.}/\rho_0$ versus $L$ at $T=0$.
}
\end{figure}

For completeness, we report  on
the baryon density in the subcritical liquid phase, $n_B^\mathrm{s.l.}$,
at $T=0$.
This can be estimated quickly because only the zero mode is populated therefore
we read its value from Eq.~\eqref{eq:bnm5}, namely
\begin{equation}
n_B^\mathrm{s.l.} = \frac{2N_c N_f}{3L^3};\label{eq:nbsl}
\end{equation}
this amounts to put $2N_c N_f$ quarks in the volume $L^3$.  
The results are collected in Fig.~\ref{Fig:nbsl} in which we show $n_B^\mathrm{s.l.}/\rho_0$ versus $L$,
where $\rho_0$ is the nuclear saturation density. In particular, $n_B\approx \rho_0$ for $L\approx 3$ fm.

\begin{figure}[t!]
\begin{center}
\includegraphics[width=0.45\textwidth]{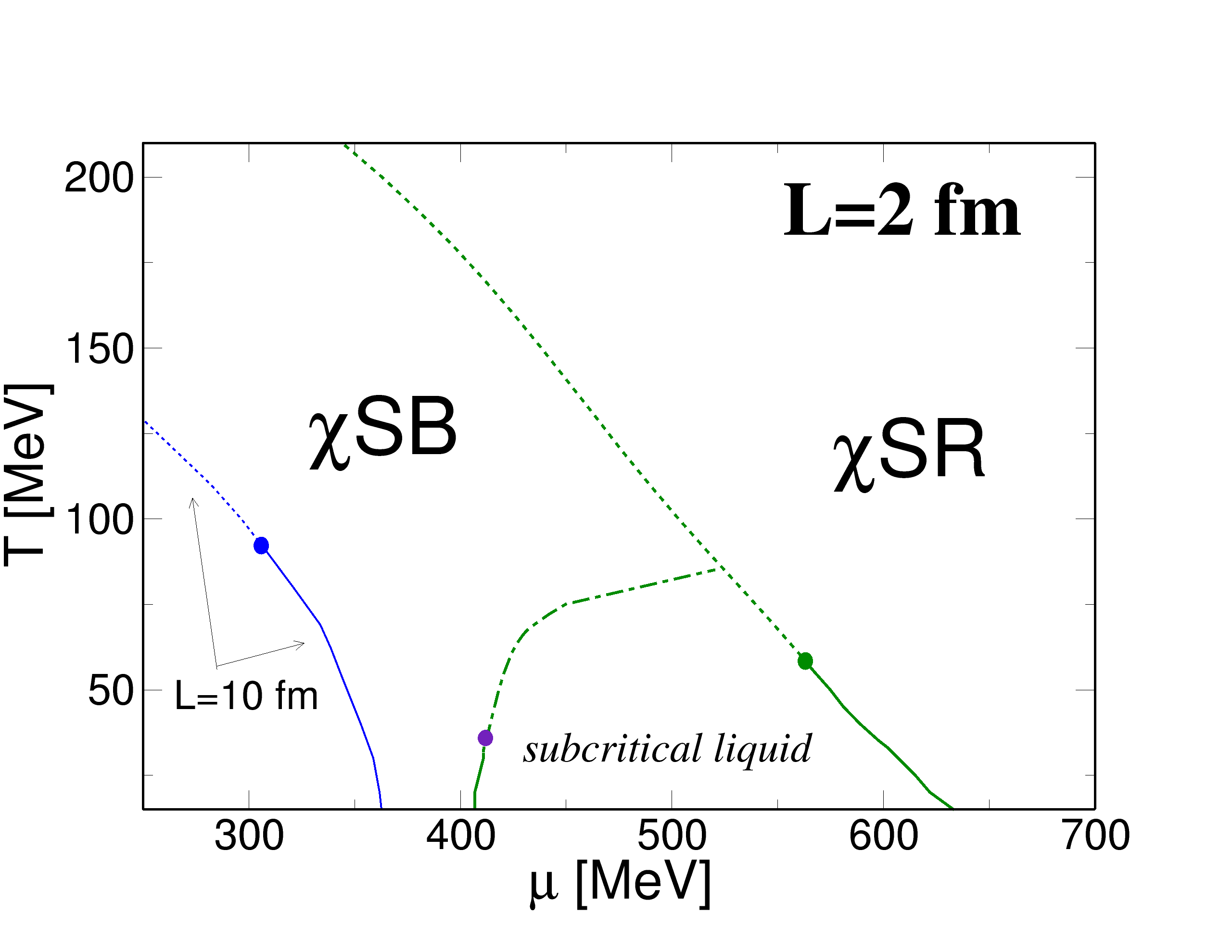}
\end{center}
\caption{\label{Fig:pdl2} Critical lines for $L=2$ fm. 
Dotted and dot-dashed lines correspond to smooth crossovers and solid lines to first order phase transitions.
The green dot denotes the critical endpoint of the chiral phase transition.
The indigo dot corresponds to the critical endpoint for the liquid-gas-like transition to the subcritical liquid phase.
$\chi$SR and $\chi$SB denote the regions in which chiral symmetry is restored and broken respectively.
We have shown by blue lines the critical lines for $L=10$ fm for comparison.}
\end{figure}  

The results discussed in this section can be summarized in the form of a phase diagram in the $\mu-T$ plane.
In Fig.~\ref{Fig:pdl2} we plot the transition lines for the case $L=2$ fm, and for comparison we also show a portion of the
critical lines at $L=10$ fm that correspond to the continuum limit.
The regions denoted with $\chi$SR and $\chi$SB denote the portions of the phase diagram 
in which chiral symmetry is restored and broken respectively. 
The dots denote critical endpoints. In the figure we focus on the subcritical region phase
that appears as an intermediate phase between $\chi$SB and $\chi$SR phases.
Comparing the critical lines for $L=2$ fm and $L=10$ fm the catalysis of symmetry breaking is evident. 
We also notice that the critical endpoint for chiral symmetry restoration moves towards higher values of $\mu$
and lower $T$ with the lowering of $L$.

We have verified the stability of our results by changing the number of colors:
in particular, for $N_c=2$ the picture is unchanged. Since QCD with $N_c=2$
and finite $\mu$ can be simulated on the lattice, the predictions of this article can be tested by means
of first principle calculations.

\section{Chiral phase transition at high temperature}
The chiral  phase transition at finite temperature and low $\mu$ has been more studied in the literature,
therefore we limit ourselves to present a few results and compare them with those of
other effective models. In particular, our results agree with those of \cite{Xu:2019gia,Xu:2019kzy}
where the NJL model with PV regulators has been used.

\subsection{Numerical computation of $T_c$ versus $L$}

\begin{figure}[t!]
\begin{center}
\includegraphics[width=0.45\textwidth]{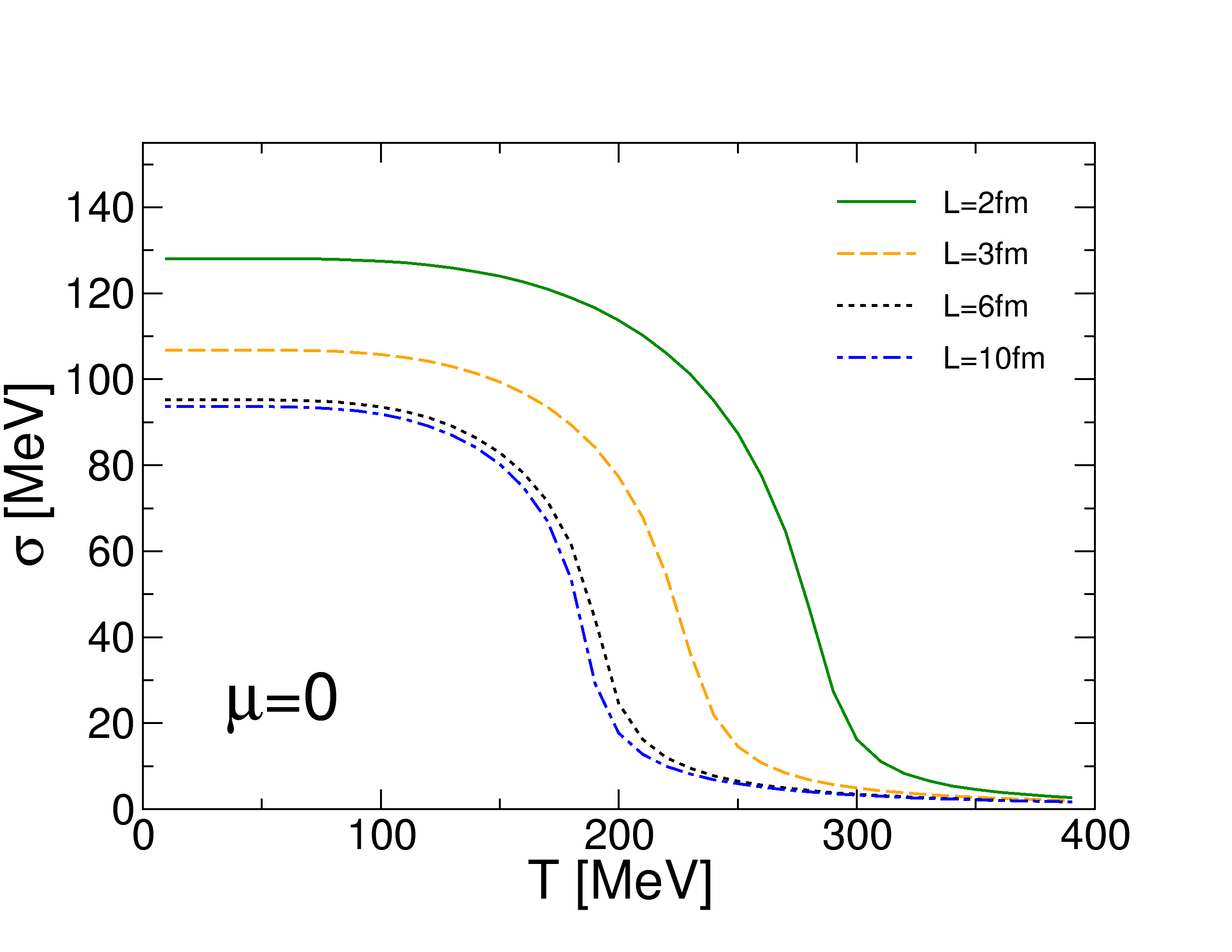}\\
\includegraphics[width=0.45\textwidth]{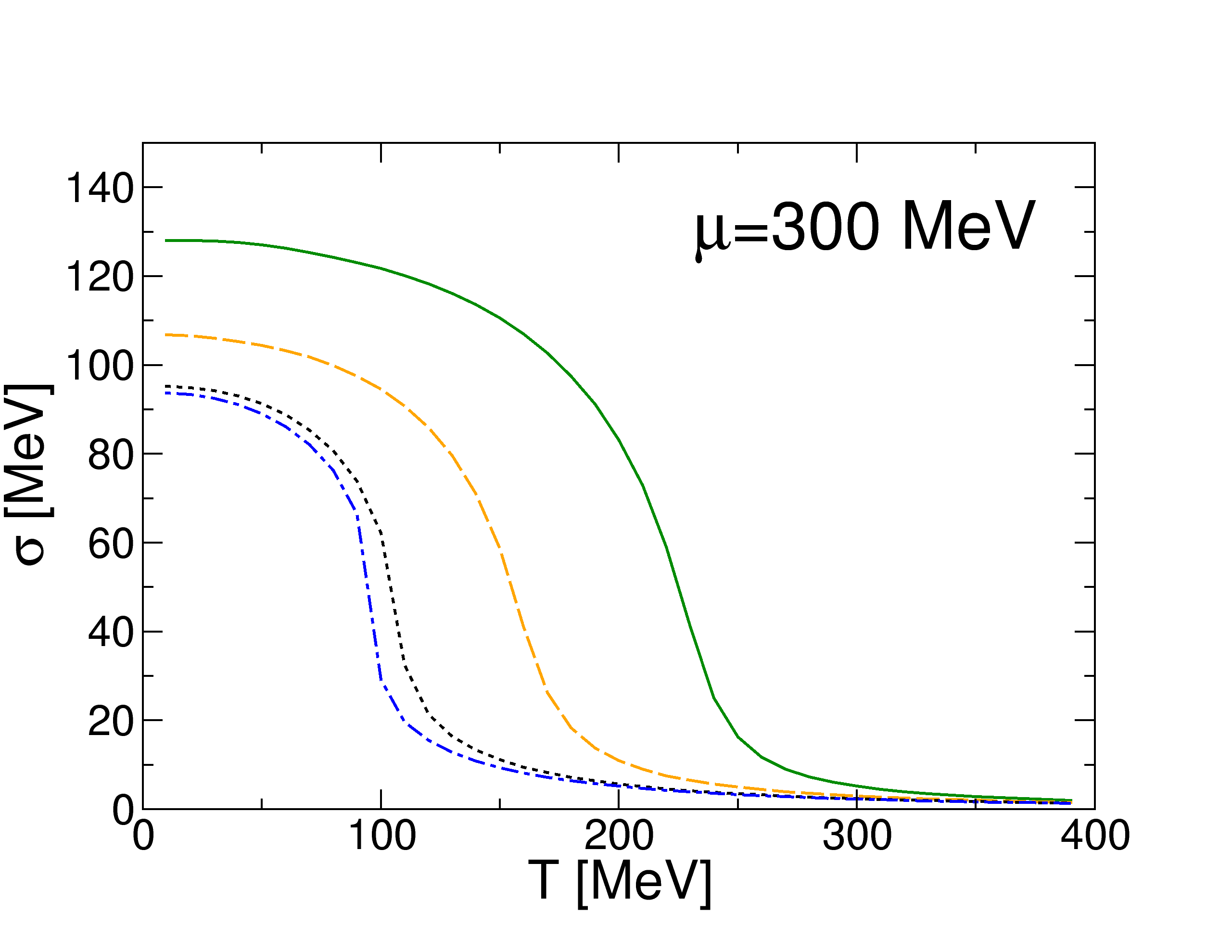}
\end{center}
\caption{\label{Fig_mm}Condensate $\sigma$ versus temperature, for several values of $L$
and two representative values of $\mu$.
}
\end{figure}

In Fig.~\ref{Fig_mm} we plot the condensate, $\sigma$, versus temperature, for $\mu=0$ (upper panel) and $\mu=300$ MeV
(lower panel) and several values of $L$. 
We can define a pseudo-critical temperature, $T_c$, by looking at the location of the maximum variation of
$d\sigma/d\beta$. At $\mu=0$ the condensate increases with $1/L$ and this pattern remains stable in the whole
temperature range examined. We conclude that our picture is consistent with the catalysis of chiral symmetry breaking
induced by lowering $L$. The catalysis remains also for higher values of $\mu$, see the lower panel 
of Fig.~\ref{Fig_mm}, in agreement with the results presented in section~\ref{Sec:llp}.

\begin{figure}[t!]
\begin{center}
\includegraphics[width=0.45\textwidth]{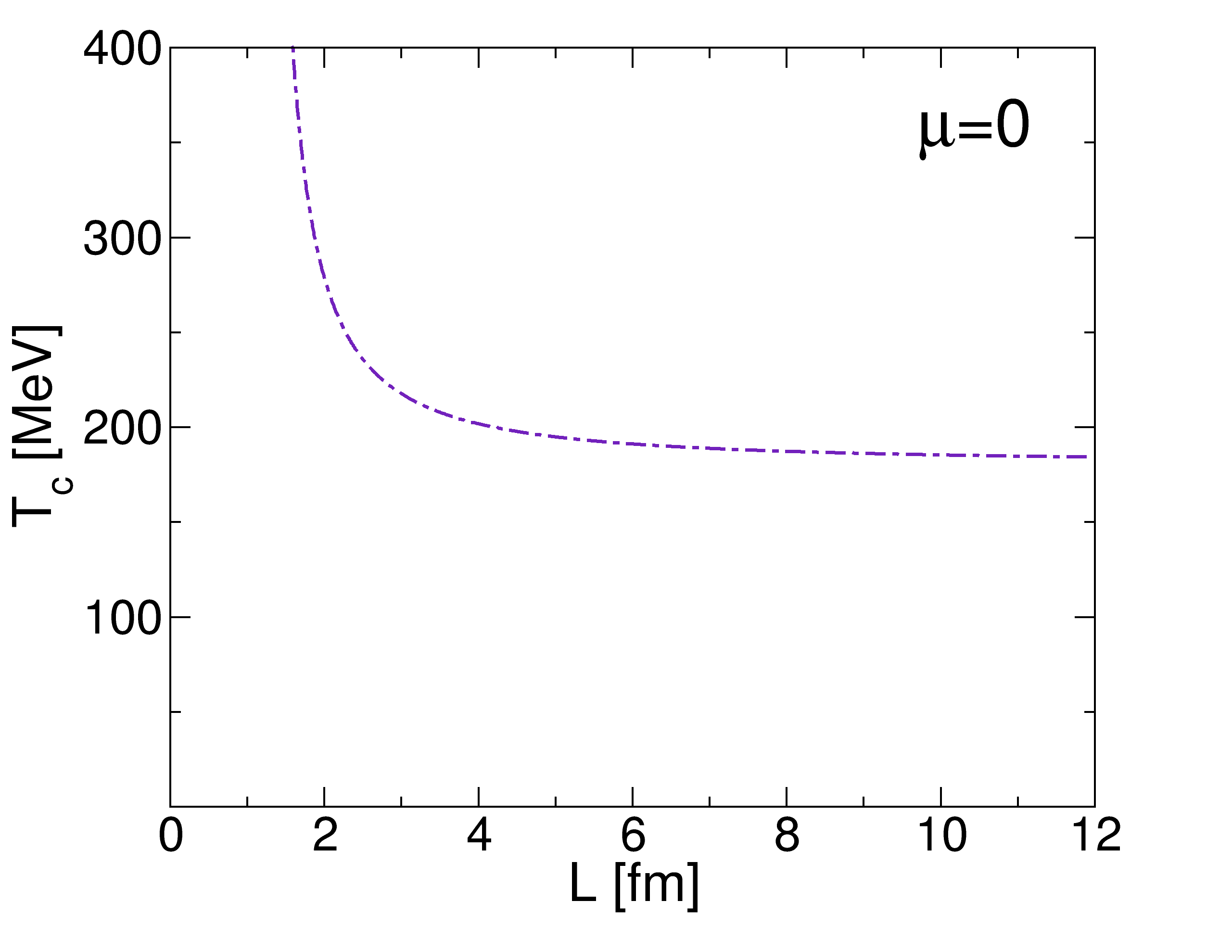}
\end{center}
\caption{\label{Fig:tctc}Critical temperature for chiral symmetry restoration versus size at $\mu=0$.
The line represents a crossover for any value of $L$.
}
\end{figure} 
 
We show $T_c$  versus $L$ at $\mu=0$ in Fig.~\ref{Fig:tctc}.
The behavior for other values of $\mu$ can be easily guessed from the results that we have shown before.
In the infinite volume limit $T_c\approx 180$ MeV. The catalysis of chiral symmetry breaking is
clear in Fig.~\ref{Fig:tctc}. For example, for $L=3$ fm the increase of critical temperature is $\approx 23\%$,
while it becomes $\approx 55\%$ for $L=2$ fm.  The qualitative behavior of $T_c$ is in agreement 
with previous studies within the QM model~\cite{Palhares:2009tf,Magdy:2019frj}
as well as the NJL model with periodic boundary conditions \cite{Xu:2019gia,Xu:2019kzy,Wang:2018ovx}.

\subsection{Critical temperature from the Ginzburg-Landau potential}
In this subsection we clarify the role of the zero mode 
on the behavior of $T_c$ versus $L$. We do this in the chiral limit, $h=0$, and at $\mu=0$, 
to make the discussion more transparent. 
In these conditions the chiral transition is of the second order and the critical temperature is given by the zero
of the coefficient $\alpha_2$ of the Ginzburg-Landau potential,
\begin{equation}
\alpha_2 = \left.\frac{\partial^2\Omega}{\partial\sigma^2}\right|_{\sigma=0},
\end{equation}
where  $\Omega$ is given by Eq.~\eqref{eq:tot_om_fs}. 
The coefficient $\alpha_2$ corresponds to the curvature of $\Omega$ at $\sigma=0$.
We feel that this discussion is necessary because in the literature some confusion arises when different models 
are compared to each other.
We keep the parameters of the classical potential,
$v$ and $\lambda$, unchanged by $L$. This is natural within the renormalization scheme 
that we have adopted in this article, because the finite size corrections to $\Omega$
are finite and do not require any additional renormalization condition.

Starting from Eq.~\eqref{eq:tot_om_fs}  we get
\begin{eqnarray}
\alpha_2^\mathrm{QM} &=& C_2^\mathrm{QM} \nonumber\\
&&+\frac{4 N_c N_f g^2}{2\pi L^2}\sum_{n=1}^\infty r_3(n)\frac{e^{-\beta\varepsilon_n}}{\sqrt{n}(1+e^{-\beta\varepsilon_n})}
\nonumber\\
&&
-\frac{4 N_c N_f }{ L^3 }\frac{g^2}{4T},\label{eq:a2QM}
\end{eqnarray} 
where we have put $\varepsilon_n = 2\pi\sqrt{n}/L$,
and $C_2^\mathrm{QM}$ denotes the curvature of $\Omega$ at $T=0$ and $\sigma=0$,
\begin{eqnarray}
C_2^\mathrm{QM} &=& -v^2\lambda + g^2\delta v \nonumber\\
&&-\frac{2 N_c N_f g^2}{L^3}
\sum_{n=1}^\infty r_3(n)\sum_{j=0}^3\frac{c_j}{\sqrt{j\xi^2 + 4\pi^2n/L^2}}.\nonumber\\
&&\label{eq:C2qm}
\end{eqnarray} 
We remind that $m_\sigma^2=2v^2\lambda$ corresponds to the $\sigma-$meson mass
in the vacuum and in the infinite volume limit.
The last addendum on the right hand side of 
Eq.~\eqref{eq:a2QM} is the zero mode contribution to $\alpha_2$,
while the summation represents the contribution of the higher modes.
The zero mode contribution is negative while the sum over the higher modes is positive:
while thermal fluctuations increase $\alpha_2$ making the broken phase less stable,
the zero mode lowers $\alpha_2$ causing the broken phase to be more stable.
A numerical inspection shows that $C_2^\mathrm{QM}$ is quite insensitive of $L$
because it is dominated by the classical contribution. Moreover $C_2^\mathrm{QM}<0$.

Lowering $L$, the zero mode contribution grows up in magnitude, therefore it is necessary to increase
$T_c$ to get a positive contribution from the higher modes that overcomes both the zero mode and $C_2^\mathrm{QM}$
to satisfy $\alpha_2^\mathrm{QM}(T_c)=0$. Thus $T_c$ increases by lowering $L$.

If we removed the zero mode from Eq.~\eqref{eq:a2QM}, for example by imposing antiperiodic boundary condtions
or an infrared cutoff,
we would be left with
\begin{eqnarray}
\alpha_{2,\mathrm{nzm}}^\mathrm{QM} &=& C_2^\mathrm{QM}   \nonumber\\
&&+\frac{4 N_c N_f g^2}{2\pi L^2}\sum_{n=1}^\infty r_3(n)\frac{e^{-\beta\varepsilon_n}}{\sqrt{n}(1+e^{-\beta\varepsilon_n})}.
\label{eq:a2QMapbc}
\end{eqnarray}
Even in this case, the requirement $\alpha_{2,\mathrm{nzm}}^\mathrm{QM}(T_c)=0$
implies that
lowering $L$ has to be balanced by the increase of $T_c$,
because $ C_2^\mathrm{QM}<0$ and almost insensitive to $L$.
Thus in the QM model even without the zero mode,  $T_c$ has to increase with  $1/L$,
in agreement with~\cite{Palhares:2009tf,Magdy:2019frj}.
The arguments above would apply also if we had not implemented
renormalization and had regularized the divergent quark loop via an effective cutoff: in this case,
as long as $-v^2\lambda <0$, $C_2^\mathrm{QM}<0$ for any $L$.

Summarizing, we have shown that within the QM model in the chiral limit and at $\mu=0$,
$T_c$ increases with $1/L$ regardless of the presence of the zero mode in the spectrum or not.
This happens because $C_2^\mathrm{QM}<0$ for any $L$.
This behavior of $T_c$ and its explanation has not been stressed enough in the literature.

The increase of $T_c$ with $1/L$ when periodic boundary conditions are implemented is in agreement with
the results of the NJL model, see for example \cite{Xu:2019gia,Wang:2018ovx}. On the other hand,
when antiperiodic boundary conditions are implemented within the NJL model, $T_c$ is found to decrease with $1/L$.
This is in qualitative disagreement with the QM model and needs to be clarified. 
The very reason of the different behavior cannot be traced back to the lack of the zero mode only:
after all, in the QM model $T_c$ increases with $1/L$ even when the zero mode is removed from the spectrum.
In fact, a difference between the QM and NJL models is the curvature of the potential
at $\sigma=0$ and $T=0$, due to the different classical potentials in the two models.
In the QM model $C_2^\mathrm{QM}<0$.
On the other hand, in the NJL model there are no mesons at the tree level and the classical potential
is merely the mean field term $\sigma^2/4G$.  
The divergent quark loop is necessary to make $\alpha_2$ negative and break chiral symmetry:
 while this is enough to guarantee a negative curvature in the infinite volume limit,
it is not guaranteed that it remains negative for any $L$. 

To keep the treatment simple, we use the common hard cutoff scheme for the NJL model;
in fact, the results we find here agree with those obtained within
PV regularization \cite{Xu:2019gia}.
Using a  cutoff $\Lambda$ the curvature at $T=0$ in NJL is
\begin{equation}
C_2^\mathrm{NJL} = \frac{1}{2G}  -\frac{N_c N_f}{\pi L^2}\sum_{n=1}^a \frac{r_3(n)}{\sqrt{n}},
\end{equation}
where $a=\Lambda^2 L^2/4\pi^2$ and $G$ is the NJL coupling. The contribution of the thermal fluctuations to $\alpha_2$ in NJL
is formally equivalent to that of QM model and is not repeated here.
We notice that  $C_2^\mathrm{NJL}$
becomes positive for small enough $L$ because the quark loop shrinks. 
When the zero mode is removed from the spectrum, 
the quark loop in $C_2^\mathrm{NJL}$ is the only source for a negative $\alpha_2$:
since lowering $L$ shrinks the loop, the contribution of the excited states at $T_c$ has to be lowered to get a vanishing $\alpha_2$
which implies that $T_c$ decreases when $L$ decreases.

The message of this section is that the zero mode alone is not enough to explain the behavior of $T_c$ versus $L$
in chiral models: the difference between QM and NJL appears even when the zero mode is absent in the
spectrum of both models. The curvature of the potential is another necessary ingredient for $T_c$ and it is precisely the
different curvature that leads to different predictions of $T_c$ versus $L$ in the two models.

We remark that if we had assumed 
a dependence of the classical potential of $L$, the behavior of $T_c$ versus $L$ might have been more difficult to predict
within the Ginzburg-Landau coefficient because $T_c(L)$ would have depended also on 
the additional functions $v=v(L)$ and $\lambda=\lambda(L)$.

\section{Summary and Conclusions}
We have studied the effect of periodic boundary conditions on chiral symmetry breaking and its restoration in QCD.
As an effective model of the effective potential for the quark condensate, we have used the quark-meson model 
which couples quarks to background meson fields.
We have implemented periodic boundary conditions on the effective potential for a cubic box of size $L^3$,
then we have performed the renormalization of the divergent vacuum term in the box;
we have computed the behavior of the condensate at finite temperature, $T$, and quark chemical potential, $\mu$.
For the implementation of the renormalization conditions at finite $L$ we have 
adopted the Pauli-Villars regulators as in \cite{Xu:2019gia}, that are enough to cancel the divergent contributions
in the infinite volume as well as at finite $L$.

The most interesting effects happen for the chiral phase transition at small temperature and finite chemical potential.
We have found that for $L\lesssim 5$ fm, increasing $\mu$ up to a critical value, $\mu_1$, 
results in a jump of the condensate
to lower but finite values. This jump is due to the population of the zero mode.
The contribution of the zero mode at such moderate size is not very strong, therefore its excitation is not enough
to restore chiral symmetry. Increasing $\mu$ to higher values, the first mode is excited eventually
and chiral symmetry is restored at $\mu=\mu_2$.

We have suggested a similitude between the jump of the condensate at $\mu=\mu_1$ and a liquid-gas phase transition.
In fact, chiral symmetry is not restored at $\mu_1$,
therefore symmetries are broken in the same way at low and intermediate $\mu$.
Moreover, the quark number density at the first jump has a net increase.
Both these aspects are common to the liquid-gas phase transition.
We have further characterized the jump of the condensate at $\mu_1$ by means of the correlation length
of the fluctuations of the condensate, that are carried by the $\sigma-$meson; in particular,
we have identified $\lambda=1/M_\sigma$ with $\lambda$ the correlation length and $M_\sigma$ the in-medium
mass of the $\sigma-$meson. We have found that at low temperature and $\mu=\mu_1$ the correlation length
increases then stays constant up to $\mu=\mu_2$ where chiral symmetry is restored. Increasing temperature
brings the system close to criticality and this is confirmed by the increase of $\lambda$.

We name the intermediate phase as {\em subcritical liquid phase} because even though
the system is not critical in the whole $(\mu-T)$ window, the correlation domains in this phase
are larger than those in the hadron gas and quark matter phases, respectively at small and large $\mu$,
as if the system was approaching a critical point; in addition to this, baryon density is finite due to the occupation
of the zero mode, and symmetries are broken in the same way of the hadron phase, as it would happen in the gas-to-liquid
transition.

Overall, we have found that lowering $L$  and imposing periodic boundary conditions
catalyzes the spontaneous breaking of chiral symmetry. This has been understood as  the result of the excitation 
of the zero mode at intermediate values of $\mu$, that lowers a bit the value of the condensate without restoring chiral symmetry, 
and the need of a large $\mu$ to excite the first mode
that leads to the definitive lowering of the condensate; the smaller $L$ the larger is the value of $\mu$ needed to excite the first mode,
thus leading to the catalysis of chiral symmetry breaking.

We have completed the study by computing 
the critical temperature,
$T_c(L)$, versus $L$. We have found that $T_c$ decreases with $L$ 
thus supporting the catalysis of chiral symmetry breaking 
found in previous studies where periodic boundary conditions, or effective infrared
cutoffs in the QM model, have been implemented \cite{Xu:2019gia,Palhares:2009tf,Wang:2018ovx,Magdy:2019frj}. 

There are several ways to continue the work presented here. A straightforward extension of the work is the 
inclusion of meson fluctuations on the same line of \cite{Castorina:2020vbh}.
It would be interesting to include the possibility of inhomogeneous condensates
\cite{Abuki:2018iqp,Takeda:2018ldi,Abuki:2013pla,
Abuki:2013vwa,Buballa:2020xaa,Carignano:2019ivp,Buballa:2018hux,Carignano:2017meb,Nickel:2009ke,Nickel:2009wj}.
The inclusion of the Polyakov loop  to take trace of confinement-deconfinement phase transition via a collective field
would also be possible \cite{Fukushima:2003fw,Ratti:2005jh,Abuki:2008nm}. 
Finally, it is of a certain interest to analyze the thermodynamic geometry
\cite{Weinhold:1975get,Weinhold:1975gtii,
Ruppeiner:1979trg,Ruppeiner:1983ntp,Ruppeiner:1983tcf,Ruppeiner:1985tcv,Ruppeiner:1995rgf,Ruppeiner:1998rgc,
 Castorina:2019jzw,Castorina:2020vbh,Zhang:2019neb} 
of the effective models of chiral symmetry breaking with finite size.
We plan to report on these topics in the near future.
We have also investigated the stability of our results by changing the number of colors:
in particular, we have verified that for $N_c=2$ the picture is unchanged. Since QCD with $N_c=2$
and finite $\mu$ can be simulated on the lattice, the predictions of this article can be tested by means
of first principle calculations.

\section*{ACKNOWLEDGEMENTS}
M. R. acknowledges John Petrucci for inspiration.
The work of the authors is supported by the National Science Foundation of China (Grants No.11805087 and No. 11875153).

\appendix

\end{document}